\documentclass[twocolumn,letterpaper,aps,prl,longbibliography,superscriptaddress,
floatfix]{revtex4-1}

\usepackage{multirow}
\usepackage{graphicx}
\usepackage{dcolumn}
\usepackage{bm}

\usepackage[mathlines]{lineno}
\usepackage{array}
\newcolumntype{L}{>{\centering \arraybackslash}m{2cm}}

\begin{document}

\preprint{APS/123-QED}

\title{Transverse single-spin asymmetry for very forward neutral pion 
production in polarized $p+p$ collisions at $\sqrt{s} = 510$ GeV}

\newcommand{\korea}{Korea University, Seoul 02841, Korea}
\newcommand{\riken}{RIKEN Nishina Center for Accelerator-Based Science, Wako, Saitama 351-0198, Japan}
\newcommand{\florence}{University of Florence, Florence, Italy}
\newcommand{\florenceinfn}{INFN Section of Florence, Florence, Italy}
\newcommand{\catania}{University of Catania, Catania, Italy}
\newcommand{\cataniainfn}{INFN Section of Catania, Catania, Italy}
\newcommand{\icrr}{Institute for Cosmic Ray Research, University of Tokyo, 
Kashiwa, Chiba, Japan}
\newcommand{\jaea}{Advanced Science Research Center, Japan Atomic Energy 
Agency, 2-4 Shirakata Shirane, Tokai-mura, Naka-gun, Ibaraki-ken 319-1195, Japan}
\newcommand{\nagoya}{Graduate school of Science, Nagoya University, Japan}
\newcommand{\nagoyaieee}{Institute for Space-Earth Environmental Research, 
Nagoya University, Nagoya, Japan}
\newcommand{\nagoyakmi}{Kobayashi-Maskawa Institute for the Origin of Particles 
and the Universe, Nagoya University, Nagoya, Japan}
\newcommand{\rikjrbrc}{RIKEN BNL Research Center, Brookhaven National Laboratory, Upton, New York 11973-5000, USA}
\newcommand{\seoulnat}{Department of Physics and Astronomy, Seoul National University, Seoul 151-742, Korea}
\newcommand{\tokushima}{Tokushima University, Tokushima, Japan}
\newcommand{\waseda}{RISE, Waseda University, Shinjuku, Tokyo, Japan}
\newcommand{\shibaura}{Shibaura Institue of Technology, College of Systems Engineering and Science, 
Saitama, Minuma-ku, Japan}
\newcommand{\bnl}{Brookhaven National Laboratory, Upton, New York 11973}

\affiliation{\korea}
\affiliation{\riken}
\affiliation{\florence}
\affiliation{\florenceinfn}
\affiliation{\rikjrbrc}
\affiliation{\nagoyaieee}
\affiliation{\nagoyakmi}
\affiliation{\shibaura}
\affiliation{\nagoya}
\affiliation{\seoulnat}
\affiliation{\icrr}
\affiliation{\tokushima}
\affiliation{\jaea}
\affiliation{\waseda}
\affiliation{\catania}
\affiliation{\cataniainfn}
\affiliation{\bnl}

\author{M.H.~Kim} \affiliation{\korea} \affiliation{\riken}
\author{O.~Adriani} \affiliation{\florence} \affiliation{\florenceinfn}
\author{E.~Berti} \affiliation{\florence} \affiliation{\florenceinfn}
\author{L.~Bonechi} \affiliation{\florenceinfn}
\author{R.~D'Alessandro} \affiliation{\florence} \affiliation{\florenceinfn}
\author{Y.~Goto} \affiliation{\riken} \affiliation{\rikjrbrc}
\author{B.~Hong} \affiliation{\korea}
\author{Y.~Itow} \affiliation{\nagoyaieee} \affiliation{\nagoyakmi}
\author{K.~Kasahara} \affiliation{\shibaura}
\author{H.~Menjo} \affiliation{\nagoya}
\author{I.~Nakagawa} \affiliation{\riken} \affiliation{\rikjrbrc}
\author{J.S.~Park} \affiliation{\riken} \affiliation{\seoulnat} 
\author{T.~Sako} \affiliation{\icrr}
\author{N.~Sakurai} \affiliation{\tokushima}
\author{K.~Sato} \affiliation{\nagoyaieee}
\author{R.~Seidl} \affiliation{\riken} \affiliation{\rikjrbrc}
\author{K.~Tanida} \affiliation{\jaea}
\author{S.~Torii} \affiliation{\waseda}
\author{A.~Tricomi} \affiliation{\catania} \affiliation{\cataniainfn}
\author{M.~Ueno} \affiliation{\nagoyaieee}
\author{Y.~Makino} \affiliation{\nagoyaieee}
\author{Q.D.~Zhou} \altaffiliation[Present address: ]{Institute of Particle and Nuclear Studies, High Energy Accelerator Research Organization (KEK), Tsukuba, Japan} \affiliation{\nagoyaieee}  

\collaboration{RHICf Collaboration} \noaffiliation

\author{J. H. Lee} \affiliation{\bnl}
\author{T. Ljubicic} \affiliation{\bnl}
\author{A. Ogawa} \affiliation{\bnl}

\date{\today}


\begin{abstract}

Transverse single-spin asymmetries of very forward neutral pions generated
in polarized $p + p$ collisions 
allow us to understand the production mechanism in terms of
perturbative and non-perturbative strong interactions. 
During 2017 the RHICf Collaboration installed an electromagnetic calorimeter in 
the zero-degree region of
the STAR detector at the Relativistic Heavy Ion Collider (RHIC) 
and measured neutral pions produced at pseudorapidity larger than 6
in polarized $p$+$p$ collisions at $\sqrt{s}$ = 510 GeV.
The large non-zero asymmetries increasing both in longitudinal momentum fraction $x_{F}$ and transverse momentum $p_{T}$ have been observed at low transverse momentum $p_{T} < 1$ GeV/$c$ for the first time at this collision energy. 
The asymmetries show an approximate $x_{F}$ scaling 
in the $p_{T}$ region where non-perturbative processes are
expected to dominate. 
A non-negligible contribution from soft processes may be necessary to explain the
nonzero neutral pion asymmetries.

\end{abstract}

\maketitle


Although the largest fraction of energy in high-energy hadronic collisions is concentrated in the 
very forward region, the reaction mechanism there is not well 
understood, yet. 
RHIC has an advantage to study the production mechanism via the 
transverse single-spin asymmetry ($A_N$) of neutral 
particles in transversely polarized $p + p$ collisions.
$A_N$ is defined by 
$(d\sigma_{\rm{Left}} - d\sigma_{\rm{Right}})/(d\sigma_{\rm{Left}} + d\sigma_{\rm{Right}})$ 
or the corresponding azimuthal angular modulation 
where $\sigma_{\rm{Left (Right)}}$ designates the particle production
cross sections in the left (right) side of the polarization direction of the proton beam.
Large values of $A_N$ in hadron production have been measured in the forward 
pseudorapidity ($\eta$) region, 1 $< \eta <$ 4, in a wide range of collision energies
~\cite{Adare:2013ekj,Abelev:2008af,Adams:1991rw,Bonner:1988rv,Klem:1976ui}.

These results have been explained by Transverse Momentum 
Dependent (TMD)~\cite{Sivers:1989cc,Collins:1992kk} and higher-twist functions 
~\cite{Qiu:1991wg,Eguchi:2006mc,Kanazawa:2014dca} in an initial or 
final state effect combined with the transverse motions of quarks and gluons. 
The TMD functions are used in describing Drell-Yan or weak boson production 
where the transverse momentum $(p_{T})$ and 
momentum transfer $(Q^{2})$ scales are observed.
On the other hand, 
the higher-twist functions are used in inclusive hadron, photon, or jet production
processes where only a single scale in $p_T$ is observed.

Although large asymmetries could be explained by hard processes, 
recent measurements additionally suggest that they may originate from soft processes
such as diffractive scattering. 
The AnDY experiment reported small 
$A_N$ values in forward jet production, 
compared to that of forward hadron production~\cite{Bland:2013pkt}. 
The difference can be explained not only by the mixture and cancellation of
$u$ and $d$-quark jets, but potentially also by diffractive effects contributing
to the hadron asymmetries.

The STAR experiment also reported a multiplicity dependence of $A_N$ 
for $\pi^{0}$ with the number of detected photons~\cite{Mondal:2014vla}. 
It showed that the $A_N$ decreases as the event complexity increases and
jet-like events show small asymmetries. 
It poses a question whether the large $A_{N}$ values of $\pi^0$ 
are due to diffractive scattering. 
This study tries to understand the asymmetries in the region where
soft processes dominate by measuring $A_{N}$ of $\pi^{0}$ at very
forward rapidities and small $p_{T}$. 

\begin{figure}[htb]
\centerline{%
\includegraphics[width=.95\hsize]{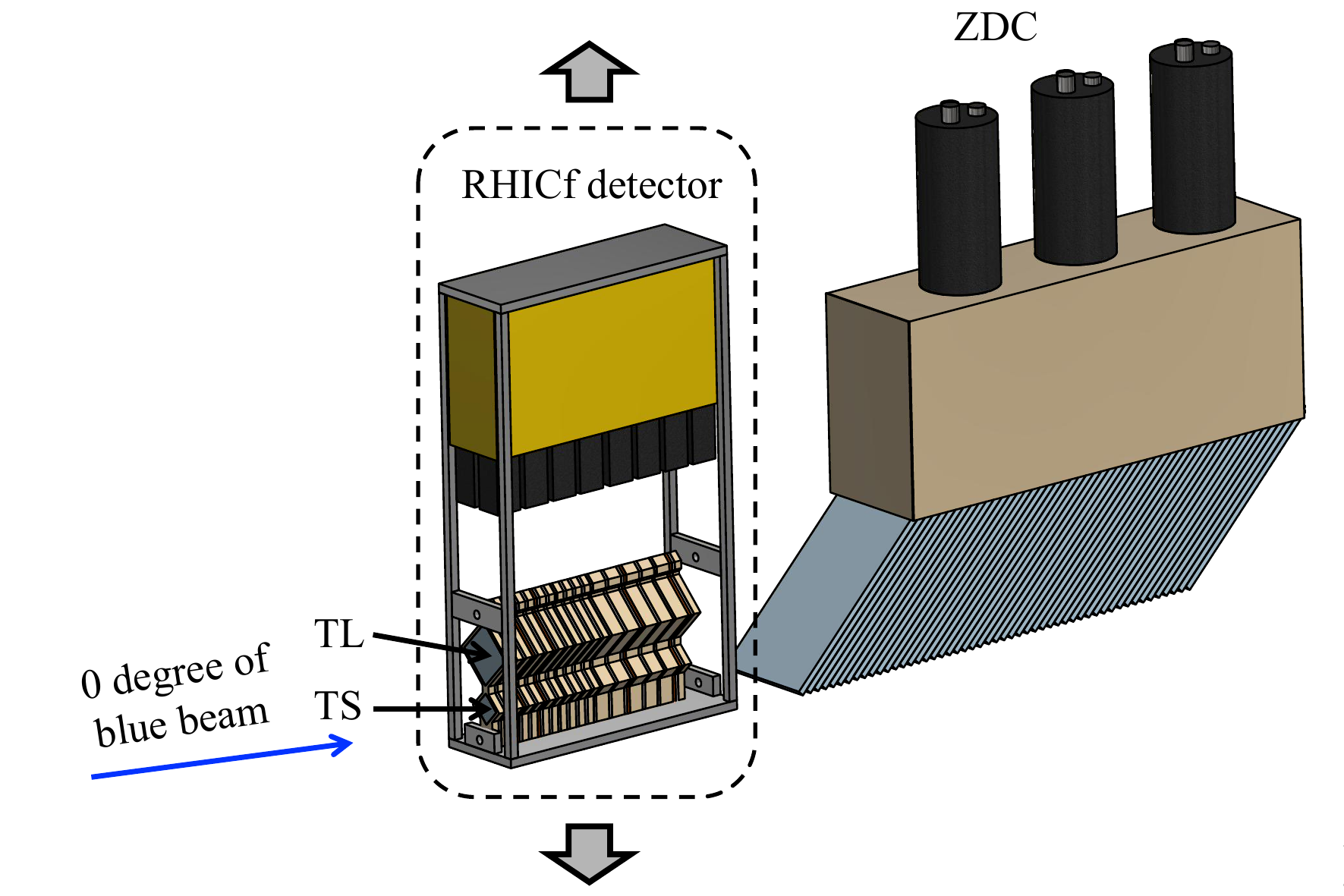}}
\caption{Schematic drawing of the RHICf detector installed in front of the ZDC with the direction of the blue beam. 
We moved the detector vertically to cover $p_{T}$ from 0.0 to 1.0 GeV/$c$.}
\label{fig:setup}
\end{figure}

We installed the former LHCf Arm1 detector~\cite{Adriani:2008zz},
now dubbed as RHICf detector,
in front of one of the
STAR Zero-Degree Calorimeters (ZDC) \cite{starzdc}, which was located 
18 m away from the beam collision point as shown in Fig. \ref{fig:setup}.
At RHIC, the proton bunches rotating clockwise are referred to as "blue beam"
and counterclockwise as "yellow beam",
and the RHICf detector is located in the downstream side of the blue beam.
The direction of 0 degree of the blue beam is also shown in Fig. \ref{fig:setup}.
The RHICf detector consists of the two position-sensitive sampling calorimeters
with square shape in the transverse plane, 
called TS (small tower, with a 20 mm size) and TL (large tower, with a 40 mm size). 
Each detector consists of 17 tungsten absorbers with a total of 44 radiation le./ngths (or 1.6 
nuclear interaction lengths), 16 sampling layers of GSO scintillators, and 4 X-Y pairs 
of the position layers with multianode PMT readout. 
Each pair of position layers is composed of 20 (for TS) or 40 (for TL) 1 mm wide GSO bars~\cite{gsobar}.
The RHICf detector has an energy resolution of 2$\sim$3$\%$ and position 
resolution of 100$\sim$150 $\mu$m for 100$\sim$250 GeV photons.
The two photons from $\pi^{0}$ decays can be detected in two different towers (Type-I) or
within one tower (Type-II)~\cite{Adriani:2015iwv}. 
For $\pi^{0}s$, the energy resolution
at energies of 100$\sim$250 GeV is 2.5$\sim$3.5$\%$ and the
$p_{T}$ resolution in $0.0 < p_{T} < 0.8$ GeV/$c$ is 3.0$\sim$4.5$\%$
for both types.

In order to cover a wide $p_{T}$ range, 
the data were taken at three vertical positions of the detector where the beam
enters (1) the center of the TL, (2) the center of the TS, and (3) 24 mm below the center of the TS.
With three configurations, 
we were able to measure $A_N$ of the very forward $\pi^0$s at RHIC in 
$0.0 < p_T <$ 1.0 GeV/$c$.
The measured longitudinal momentum fraction $(x_{F})$ region was 
larger than $0.25$.
Further limits were imposed by the shadows of the upstream horizontal bending magnet, DX,  
and the beam pipe, resulting in the accepted pseudorapidity region of $\eta > 6$.
Note that the current RHICf setup kinematically covers a lower $p_{T}$ region
than the PHENIX and STAR measurements, which is beneficial to studying the soft
quantum chromodynamics (QCD) effects.

The RHIC proton beams are usually vertically polarized.
111 of 120 bunches are filled with beam spin orientation up or down
in fixed bunch patterns.
During the RHICf operation in 2017, the direction of the beam polarization was rotated 
by 90 degrees by using the spin rotator magnets to radial polarization
in order to reach maximal sensitivity of $A_{N}$ and a maximal $p_{T}$
range covered by the RHICf detector.
The polarization ranged from 0.53 to 0.59 with a systematic uncertainty less than 0.02 \cite{rhicpol}.
The center-of-mass energy ($\sqrt{s}$) was 510 GeV 
with large $\beta^*$ value of 8 m to 
make the angular beam divergence small. As a result, the luminosity of $\sim$$10^{31}$ cm$^{-2}$s$^{-1}$ in the RHICf operation was smaller than that
in usual RHIC operation with a small $\beta^*$.

Three kinds of triggers were generally used for the neutral particle
measurement. A shower trigger with a large prescale factor of $<$ 30
was the baseline trigger. It required hits in three 
consecutive GSO sampling layers of the TS or TL tower.
The Type-I $\pi^{0}$ trigger was designed for the entire luminosity 
without any prescale factor. 
It required hits in the three consecutive layers in the upstream seven 
sampling layers of both TS and TL. 
Finally, the high-energy electromagnetic (high-EM) trigger 
was designed and optimized for the measurements of
high-energy photons and Type-II $\pi^0$s.
It required a large energy deposit in the fourth sampling layer of the TS 
or TL, and was operated with a small prescale factor of $\sim$2. 
In total, 1.1$\times10^{8}$ events and an integrated luminosity of about 700 nb$^{-1}$ 
were accumulated in four days of the dedicated RHIC operation spanning about 28 hours of data taking.


\begin{figure}[h]
\centering
\includegraphics[width=0.34\textwidth]{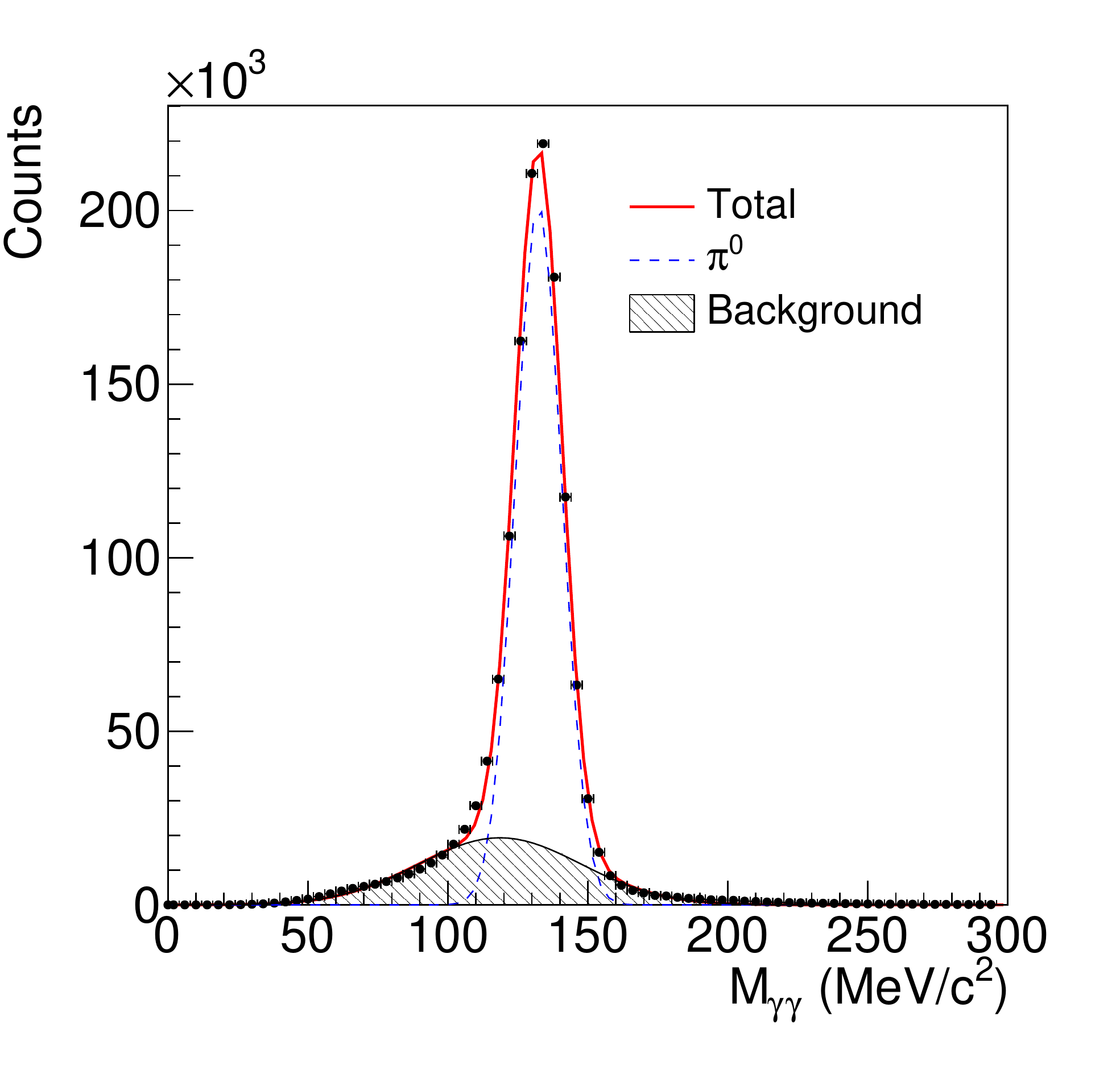}
\caption{Reconstructed two photon invariant mass distribution of Type-I in $x_{F} > 0.25$ and $0.0 < p_T <$ 1.0 GeV/$c$.} 
\label{fig:pi0}
\end{figure}

The hit positions of photons were estimated by fitting a Lorentzian-based
function to the energy deposit distribution in the GSO bars.
The photon energy was reconstructed based on the correlation between
the energy deposit in the detector and its incident energy simulated by {\sc GEANT4}~\cite{geant4}. The position dependent
light collection efficiency and shower leakage effect were also considered
for the reconstructed hit position.
See Ref.~\cite{Adriani:2015iwv, LHCf_recon} for more details on the correction procedure.

The photon events were separated from neutron background by
requiring $8 < L_{90\%} < 18$, where $L_{90\%}$ is defined by the longitudinal depth for the measured energy deposit to reach $90\%$ of
the total one.
The rejection efficiencies for neutron and photon events with the above $L_{90\%}$ 
criterion are $99\%$ and $4\%$, respectively, based on 
the {\sc GEANT4} simulation using the {\sc QGSP$_{-}$BERT 4.0} model.
The neutron contamination in the $\pi^{0}$ sample was further suppressed 
to less than $0.1\%$ by applying the two photon invariant mass cut which will 
be described later.
Due to the poor energy and position resolutions,
the photon hits in the DX magnet shadow region and the regions less than
2 mm from the detector edges were excluded in this analysis.
The $A_{N}$ distribution was analyzed as functions of $x_{F}$ and $p_{T}$ 
and the two-dimensional dependence was investigated.
The boundaries of the $x_{F}$ and $p_{T}$ bins
were determined in such a way that 
the $A_{N}$ of all bins were not 
biased by specific detector position or the types of $\pi^{0}$.

For the forward single-spin asymmetry, only the polarization of the blue beam,
the one moving towards the RHICf detector, is taken into account.
On the other hand, for the backward asymmetry, where the sign of the $x_{F}$ is reversed, 
only the polarization of the yellow beam is taken into account.
The $A_{N}$ value in each $x_{F}$ and $p_{T}$ bin was estimated by
\begin{eqnarray}
A_{N} = \frac{1}{PD_{\phi}}\Big(\frac{N_{L}-
RN_{R}}{N_{L}+RN_{R}}\Big),
\label{eq:AN}
\end{eqnarray}
where $P$ is the beam polarization, $N_{L (R)}$ is the 
number of detected $\pi^{0}$ in the left (right) side 
of the beam polarization direction, and $R$ is the luminosity ratio
of the spin orientations resulting into 
the events to right and left sides. The value of $R$, ranging from 0.958 to 0.995 was estimated using the charged
particle rates tagged by the STAR Beam Beam Counter 
(BBC)~\cite{starbbc} and Vertex Position Detector (VPD)~\cite{starvpd}.
The dilution of the asymmetry in azimuthal angle ($\phi$) of $\pi^{0}$
was corrected using the dilution factor $D_{\phi}$ defined as
\begin{eqnarray}
D_{\phi} = \sum_{i}\Big(\frac{\sin\phi_{i}}{N}\Big),
\end{eqnarray} 
where $\phi_{i}$ is the $\phi$ of $\pi^{0}$ from the beam polarization direction in the $i$-th event and 
the $N$ is the number of the detected $\pi^{0}$s.
In this analysis, only Type-I $\pi^{0}$ triggered events were used for the Type-I analysis and high-EM triggered events
for the Type-II analysis before combining the asymmetries. 
In this way, the trigger efficiencies cancel in Eq. (\ref{eq:AN}).

The obtained $A_{N}$ was corrected for the background 
contamination and the detector smearing effect. 
Figure \ref{fig:pi0} shows the reconstructed invariant mass 
distribution of the two photons and a clear $\pi^0$ peak at 135 MeV/$c^{2}$ is observed. 
Most of the background ($>$ 80$\%$) comes from accidental coincidences 
between photons from different $\pi^{0}$s.
Another major background is the combinatorial two particle background 
from direct photons, photons
from $\eta$ decays, and mis-identified neutrons.
According to the {\sc QGSJET II-04} model~\cite{qgsget}, the
distribution near the peak is well described by the superposition of the Gaussian
(for the $\pi^{0}$ peak) and the 6 th order polynomial (for the background) function.
The fitted function and their sum are shown in Fig. \ref{fig:pi0}.
The width of 3$\sigma$ around the peak position was chosen 
for the signal+background region, and the
regions beyond 5$\sigma$ to the left or right of the peak position were
chosen as the pure background regions.
The background-to-signal ratio, $N_{B}/N_{S}$, was estimated in the
signal+background region.
It decreases as $x_{F}$ increases 
($54\%$ at the lowest and $2\%$ at the highest $x_{F}$ bin) because 
higher energy $\pi^{0}$s require narrower opening angles, which
reduces the acceptance of accidental coincidences.

The background asymmetries were subtracted by
\begin{eqnarray}
A_{N}^{S} = \Big(1+\frac{N_{B}}{N_{S}}\Big)A_{N}^{S+B} - 
\Big(\frac{N_{B}}{N_{S}}\Big)A_{N}^{B},
\end{eqnarray}
where $A_{N}^{S+B}$, $A_{N}^{S}$, and $A_{N}^{B}$ are the estimated asymmetries
in the signal+background, signal only and background only, respectively, regions. According to the Monte-Carlo simulation,
the two photon invariant mass distribution has a small tail to the
lower mass region due to the underestimated reconstruction energy in a few events.
A small fraction ($\sim$4$\%$) of real $\pi^{0}$ events in the background region contributes to this tail
and the actual $N_{B}/N_{S}$ can be smaller than the one estimated by the fitting
due to these $\pi^{0}$ events.
Because the probability of the underestimated reconstruction 
energy increases
when the photon hit approaches to the edge of the tower,
even larger area along the edge than the one applied for $\pi^{0}$
,with the width of 4 mm, was excluded for the
background estimation.
With this condition, the $\pi^{0}$ tail in the background region almost disappears ($< 0.05\%$).  
All background asymmetries were consistent with zero within the statistical 
uncertainties. 
The variation of $A_{N}$ for the $\pi^{0}$ distributions with and without the tail was considered as one of the systematic uncertainties.
The typical size of this uncertainty is $\sim$0.0003.

For additional sources of the systematic uncertainty,
the variation in the beam center position and the
smearing effect for $x_{F}$ and $p_{T}$ were considered.
The beam center is obtained by extrapolating the direction
of the blue beam to the detector as follows.
In the first method, the high energy neutron hit 
distribution was fitted by the two-dimensional Gaussian function.
In the second method, the neutron asymmetries were scanned as a function of their mean
vertical position by using the fact that
very forward neutrons have zero asymmetry 
at the vertical position of the beam center~\cite{neutronAN}.
The difference of the two determined beam centers was less than 1.3 mm 
and the corresponding systematic uncertainty was 0.0003$\sim$0.0089 depending on $x_{F}$ and $p_{T}$ which is assigned to the systematic uncertainty of $A_{N}$.

\begin{figure}[htb]
\centerline{%
\includegraphics[width=0.95\hsize]{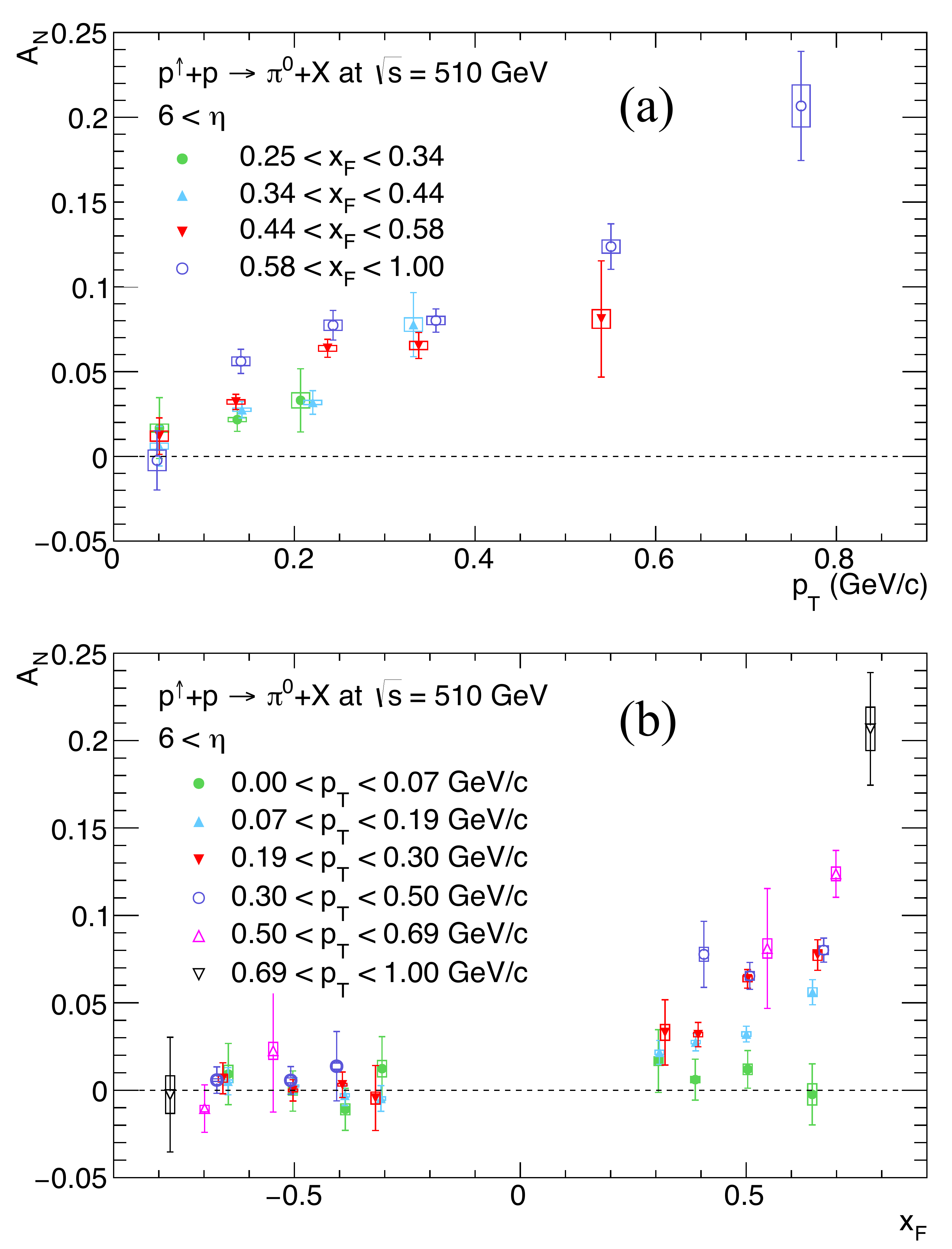}}
\caption{$A_{N}$ of the very forward $\pi^{0}$s as functions of
(a) $p_{T}$  for 
several $x_{F}$ ranges and (b) $x_{F}$ for several $p_{T}$ ranges. 
Only forward $A_{N}$ was presented in (a).
Error bars represent the statistical uncertainties, and the boxes represent
the systematic uncertainties.}
\label{fig:result}
\end{figure}

The effect of smearing due to the resolutions of $x_{F}$ and $p_{T}$ was studied in detail with {\sc GEANT4}.
The dependence of
$\pi^{0}$ asymmetries on $x_{F}$ and $p_{T}$ were artificially generated using weights.
Single $\pi^{0}$s were generated considering the detection efficiencies
matching the reconstructed energy and $p_{T}$ distributions
of the data.
The simulation was tuned to the data 
for the beam profile, detector noise, signal attenuation, and measured fluctuations
including the cross talk effect in the
GSO bars described in Ref.~\cite{lhcfJINST}. 
The data analysis code was also used in the simulation for the reconstruction of $\pi^{0}$s and the calculation of the asymmetry.
The contamination level defined by the ratio of the incorrectly and
the legitimately reconstructed events in a given ($x_{F}$, $p_{T}$) bin was
estimated in the simulation, where the reconstructed $x_{F}$ and $p_{T}$ values of the incorrect events were out of the range for the bin and the true values belong to.
The contamination level for Type-I and Type-II $\pi^0$ are all less than $35\%$, and 
more than $90\%$ of the migrated events are from
$\delta x_{F} < 0.025$ and
$\delta p_{T} < 0.035$ GeV/$c$ of the bin boundaries. 
The differences between the reconstructed and true $\langle x_{F}\rangle$, $\langle p_{T}\rangle$, and $A_{N}$ values of each bin due to 
smearing are less than 0.008, 0.009 GeV/$c$, and 0.0015, respectively, 
which are negligible.
This result is in agreement with our expectations
because the resolutions of $x_{F}$ and $p_{T}$
of the detector are much smaller than the bin sizes.

\begin{figure}[hb]
\centerline{%
\includegraphics[width=1.05\hsize]{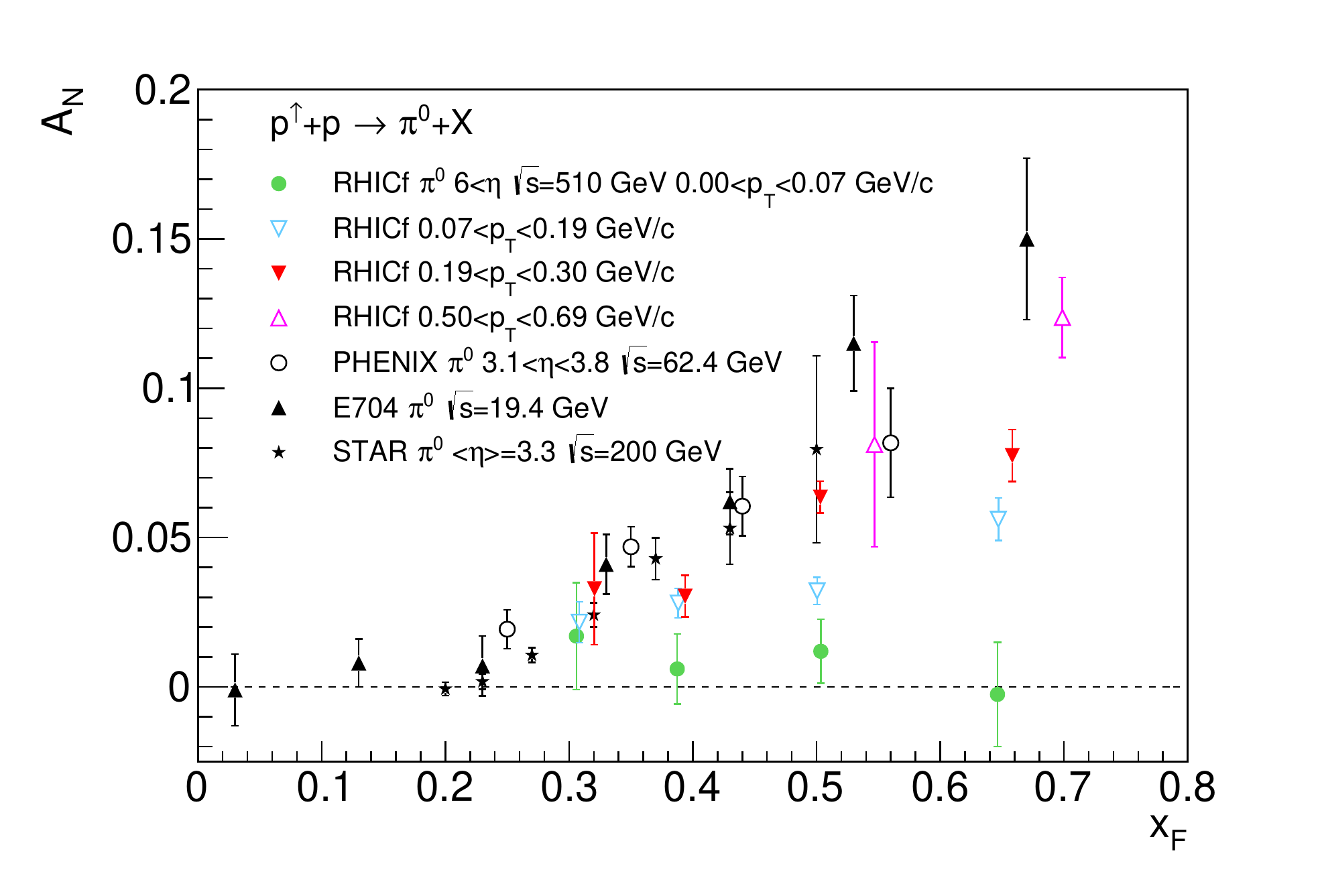}}
\caption{Comparison of the RHICf data with the previously measured $A_{N}$ of
the forward $\pi^{0}$s as a function of $x_{F}$.}
\label{fig:xFcomp}
\end{figure}

Furthermore, to find any missing systematic effects having not been considered, 
the ``bunch shuffling" analysis was performed by randomly assigning
the bunch numbers given for a fixed polarization pattern. Ideally, the calculated asymmetries and their fluctuations
after the bunch shuffling
should be centered around zero with the width of the statistical uncertainties of the asymmetries.
The bunch shuffled asymmetries were consistent with zero with comparable
fluctuations with statistical uncertainties. Therefore, we,
finally, concluded that there are no noticeable fake asymmetries introduced 
in the experiment and analysis.


\begin{table*}[ht]
\begin{center}
\begin{tabular}{c | c | c | c | c | c | c | c}
\hline \hline
\multirow{2}*{$\langle x_{F}\rangle$} & 
\multirow{2}*{$\langle p_{T}\rangle$ (GeV/$c$)} & 
\multirow{2}*{$A_{N}$} &
\multirow{2}*{Statistical uncertainty} & \multicolumn{4}{c}{Systematic uncertainty}\\
\cline{5-8} & & & & Total & Beam center & Polarization & Background\\
\hline
$-0.79$ & 0.77 & 0.0025 & 0.0328 & 0.0108 & 0.0107 & 0.0001 & 0.0009\\
$-0.70$ & 0.55 & 0.0105 & 0.0136 & 0.0018 & 0.0017 & 0.0004 & 0.0001\\
$-0.68$ & 0.37 & $-0.0058$ & 0.0076 & 0.0007 & 0.0006 & 0.0003 & 0.0000\\
$-0.66$ & 0.24 & $-0.0068$ & 0.0088 & 0.0022 & 0.0021 & 0.0002 & 0.0000\\
$-0.66$ & 0.15 & $-0.0051$ & 0.0078 & 0.0008 & 0.0007 & 0.0002 & 0.0000\\
$-0.64$ & 0.04 & $-0.0092$ & 0.0175 & 0.0050 & 0.0049 & 0.0009 & 0.0001\\
\hline
$-0.54$ & 0.53 &  $-0.0225$ & 0.0350 & 0.0048 & 0.0044 & 0.0020 & 0.0004\\
$-0.52$ & 0.34 & $-0.0056$ & 0.0079 & 0.0014 & 0.0012 & 0.0006 & 0.0004\\
$-0.50$ & 0.23 & 0.0001 & 0.0061 & 0.0007 & 0.0006 & 0.0001 & 0.0002\\
$-0.51$ & 0.14 & $-0.0025$ & 0.0047 & 0.0006 & 0.0005 & 0.0001 & 0.0003\\
$-0.50$ & 0.04 & 0.0004 & 0.0108 & 0.0023 & 0.0022 & 0.0001 & 0.0003\\
\hline
$-0.41$ & 0.33 & $-0.0137$ & 0.0198 & 0.0018 & 0.0016 & 0.0006 & 0.0001\\
$-0.39$ & 0.23 & $-0.0031$ & 0.0073 & 0.0007 & 0.0005 & 0.0004 & 0.0003\\
$-0.39$ & 0.14 & 0.0030 & 0.0058 & 0.0009 & 0.0007 & 0.0003 & 0.0004\\
$-0.38$ & 0.06 & 0.0109 & 0.0120 & 0.0031 & 0.0027 & 0.0012 & 0.0007\\
\hline
$-0.31$ & 0.21 & 0.0045 & 0.0186 & 0.0034 & 0.0033 & 0.0006 & 0.0003\\
$-0.31$ & 0.13 & 0.0047 & 0.0073 & 0.0009 & 0.0009 & 0.0003 & 0.0001\\
$-0.30$ & 0.04 & $-0.0123$ & 0.0181 & 0.0047 & 0.0045 & 0.0009 & 0.0009\\
\hline
0.30 & 0.04 & 0.0167 & 0.0179 & 0.0024 & 0.0023 & 0.0005 & 0.0004\\
0.31 & 0.13 & 0.0217 & 0.0068 & 0.0011 & 0.0010 & 0.0006 & 0.0000\\
0.31 & 0.21 & 0.0331 & 0.0186 & 0.0044 & 0.0043 & 0.0010 & 0.0002\\
\hline
0.38 & 0.06 & 0.0061 & 0.0116 & 0.0017 & 0.0015 & 0.0010 & 0.0001\\
0.39 & 0.14 & 0.0275 & 0.0049 & 0.0009 & 0.0003 & 0.0009 & 0.0002\\
0.39 & 0.23 & 0.0318 & 0.0069 & 0.0012 & 0.0006 & 0.0010 & 0.0003\\
0.41 & 0.33 & 0.0777 & 0.0189 & 0.0039 & 0.0019 & 0.0035 & 0.0002\\
\hline
0.50 & 0.04 & 0.0119 & 0.0107 & 0.0030 & 0.0029 & 0.0007 & 0.0001\\
0.51 & 0.14 & 0.0321 & 0.0045 & 0.0013 & 0.0007 & 0.0011 & 0.0000\\
0.50 & 0.23 & 0.0637 & 0.0053 & 0.0017 & 0.0006 & 0.0016 & 0.0001\\
0.52 & 0.34 & 0.0654 & 0.0076 & 0.0024 & 0.0017 & 0.0017 & 0.0002\\
0.54 & 0.53 & 0.0811 & 0.0342 & 0.0054 & 0.0040 & 0.0036 & 0.0006\\
\hline
0.64 & 0.04 & $-0.0024$ & 0.0174 & 0.0061 & 0.0060 & 0.0007 & 0.0000\\
0.66 & 0.15 & 0.0561 & 0.0071 & 0.0025 & 0.0014 & 0.0021 & 0.0001\\
0.66 & 0.24 & 0.0773 & 0.0087 & 0.0030 & 0.0018 & 0.0024 & 0.0001\\
0.68 & 0.37 & 0.0801 & 0.0068 & 0.0024 & 0.0011 & 0.0022 & 0.0002\\
0.70 & 0.55 & 0.1237 & 0.0134 & 0.0038 & 0.0012 & 0.0036 & 0.0004\\
0.79 & 0.77 & 0.2067 & 0.0321 & 0.0124 & 0.0081 & 0.0092 & 0.0019\\
\hline\hline
\end{tabular}
\caption{$A_{N}$ of very forward $\pi^{0}$s as a function of $\langle x_{F}\rangle$ 
and $\langle p_{T}\rangle$.} 
\label{tab:aspT}
\end{center}
\end{table*}

Figure~\ref{fig:result} and Table \ref{tab:aspT} summarize the 
$A_{N}$ values of very forward $\pi^{0}$s as functions of $x_{F}$ and 
$p_{T}$. 
In addition to the systematic effects described above, additional systematic uncertainties of 0.0005$\sim$0.0092 were caused by the variation of the determined beam
polarization.
Because all discussed systematic uncertainties
are independent, the quadratic sums of them are considered as the total 
systematic uncertainties. 
The uncertainties of the dilution factors are not included because its magnitude is less than 0.0001.

Figure \ref{fig:result}(a) shows that $A_{N}$ of the very forward $\pi^{0}$s increases 
with $p_{T}$ reaching about 0.2 at 
$\sim$0.8 GeV/$c$, where the production mechanism is mainly governed by non-perturbative QCD.
Figure \ref{fig:result}(b) shows that the backward $A_{N}$ distributions are consistent with zero.
Similarly, the forward $A_{N}$ is consistent with zero in 
$p_{T}$ $<$ 0.07 GeV/$c$ but it starts to increase as a function of $x_{F}$ at higher $p_{T}$. 
The comparison with the previous forward $\pi^{0}$ measurements
is depicted in Fig. \ref{fig:xFcomp}~\cite{fermipi0,phenixpi0,starpi0}.
It shows that the increasing trend of the very forward $\pi^{0}$ asymmetry is comparable
to the previous measurements at higher $p_{T}$ region from FNAL and RHIC 
that were successfully described by hard processes for
the $\pi^{0}$ production in polarized $p+p$ collisions \cite{Kanazawa:2014dca}. 
The current results are the first measurement showing the onset of the rising asymmetry at $p_{T} \lesssim 1$ GeV/$c$ at RHIC energy.
The present data with the previous STAR data~\cite{Mondal:2014vla}
raise the interesting question
on the relation between the soft and hard process contributions for $A_{N}$ of $\pi^{0}$s. The same question also applies to
the similar $x_{F}$ scaling phenomenon for the charged pion asymmetries at lower $\sqrt{s}$ of ZGS \cite{zgs}
and AGS \cite{ags} and that at higher $\sqrt{s}$ of FNAL and RHIC.
To answer this question, it is desired to investigate the same observables in the unexplored kinematic region 
between the low and high $p_{T}$ values of 0.8$\sim$2.0 GeV/$c$.

A clear non-zero $A_{N}$ in the RHICf data 
at low $p_{T}$ and the same $x_{F}$ scaling 
with the previous measurements for forward $\pi^{0}$s indicate that 
the diffractive processes may also contribute to the asymmetries at
higher $p_{T}$ where the hard processes are expected to be 
dominant.
For more detailed studies, STAR's central detectors and Roman pots~\cite{romanpot} will be helpful
to understand the mechanism
for the $\pi^{0}$ asymmetries and the relative contributions of the soft and hard processes.



In summary, the single-spin asymmetries of very forward $\pi^{0}$s have 
been measured by
the RHICf detector at the zero-degree area of the STAR detector at RHIC
in polarized $p+p$ collisions at $\sqrt{s}$ = 510 GeV.
Large $A_{N}$ values up to $\sim$0.2 were observed in the very forward region for $p_{T} < 0.8$ GeV/$c$.
The empirical $x_{F}$ scaling was also observed in $p_{T} >$ $0.19$ GeV/$c$ which is
similar to the data at higher $p_{T}$ region.


We thank the staff of the Collider-Accelerator Department at Brookhaven 
National Laboratory, the STAR Collaboration and the PHENIX Collaboration 
to support the experiment. 
We especially acknowledge the essential supports from the STAR members for
the design and the construction of the detector manipulator, installation/uninstallation,
integration of the data acquisition system, operation and management of all these 
collaborative activities.
This program is partly supported by the Japan-US Science and Technology 
Cooperation Program in High Energy Physics, JSPS KAKANHI (Nos. JP26247037
and JP18H01227), 
the joint research program of the Institute for Cosmic Ray Research (ICRR), University of 
Tokyo, and the National Research 
Foundation of Korea (Nos. 2016R1A2B2008505 and 2018R1A5A1025563).


\begin{thebibliography}{00}

\bibitem{Adare:2013ekj} 
  A.~Adare {\it et al.} (PHENIX Collaboration),
  Phys.\ Rev.\ D {\bf 90}, 012006 (2014).

\bibitem{Abelev:2008af} 
  B.~I.~Abelev {\it et al.} (STAR Collaboration),
  Phys.\ Rev.\ Lett.\  {\bf 101}, 222001 (2008).

\bibitem{Adams:1991rw} 
  D.~L.~Adams {\it et al.} (E581 and E704 Collaborations),
  Phys.\ Lett.\ B {\bf 261}, 201 (1991).

\bibitem{Bonner:1988rv} 
  B.~E.~Bonner {\it et al.},
  Phys.\ Rev.\ Lett.\  {\bf 61}, 1918 (1988).

\bibitem{Klem:1976ui} 
  R.~D.~Klem {\it et al.},
  Phys.\ Rev.\ Lett.\  {\bf 36}, 929 (1976).
\bibitem{Sivers:1989cc} 
  D.~W.~Sivers,
  Phys.\ Rev.\ D {\bf 41}, 83 (1990).
\bibitem{Collins:1992kk} 
  J.~C.~Collins,
  Nucl.\ Phys.\ B {\bf 396}, 161 (1993)



\bibitem{Qiu:1991wg} 
  J.~W.~Qiu and G.~F.~Sterman,
  Nucl.\ Phys.\ B {\bf 378}, 52 (1992).

\bibitem{Eguchi:2006mc} 
  H.~Eguchi, Y.~Koike and K.~Tanaka,
  Nucl.\ Phys.\ B {\bf 763}, 198 (2007).

\bibitem{Kanazawa:2014dca} 
  K.~Kanazawa, Y.~Koike, A.~Metz and D.~Pitonyak,
  Phys.\ Rev.\ D {\bf 89}, 111501 (2014)
\bibitem{Bland:2013pkt} 
  L.~C.~Bland {\it et al.} (AnDY Collaboration),
  Phys.\ Lett.\ B {\bf 750}, 660 (2015).

\bibitem{Mondal:2014vla} 
  M.~M.~Mondal (STAR Collaboration),
  PoS DIS {\bf 2014}, 216 (2014).
  
\bibitem{Adriani:2008zz}
O. Adriani {\it et al.} (LHCf Collaboration), Phys. Rev. D {\bf 94}, 032007 (2016).  
  
\bibitem{starzdc}
C.~Adler {\it et al.} Nucl. Intru. Meth. A {\bf 470}, 488 (2001).

\bibitem{gsobar}
T.~Suzuki {\it et al.}, JINST 8 T01007 (2013).

\bibitem{Adriani:2015iwv} 
  O.~Adriani {\it et al.} (LHCf Collaboration),
  Phys.\ Rev.\ D {\bf 94}, 032007 (2016).

\bibitem{rhicpol}
RHIC p-Carbon Measurements,\\ 
{\tt https://www.phy.bnl.gov/cnipol/rundb/}.

\bibitem{geant4}
The Geant4 Collaboration, Nucl. Instrum. Meth. \textbf{506}, 250 (2003).

\bibitem{LHCf_recon} 
O. Adriani {\it et al.} (LHCf Collaboration) Phys. Lett. B {\bf 703}, 128 (2011).

\bibitem{starbbc}
C.~A.~Whitten Jr. {\it et al.} (STAR Collaboration), The Beam-Beam Counter: A Local Polarimeter at STAR, {\tt https://www.star.bnl.gov/$\sim$eca/LocalPol}.

\bibitem{starvpd}
W. J. Llope {\it et al}., The STAR Vertex Position Detector, arXiv:1403.6855 [physics.ins-det].

\bibitem{qgsget}
S. Ostapchenko, Nucl. Phys. Proc. Suppl. {\bf 151}, 143 (2006).

\bibitem{neutronAN}
  A.~Adare {\it et al.} (PHENIX Collaboration),
  Phys.\ Rev.\ D {\bf 88}, 032006 (2013).

\bibitem{lhcfJINST}
Y. Makino {\it et al.} (LHCf Collaboration), JINST {\bf 12}, P03023 (2017).

\bibitem{fermipi0}
D. L. Adams {\it et al.} (FNAL-E581 and E704 Collaborations), Phys. Lett. B {\bf 261}, 201 (1991).

\bibitem{phenixpi0}
A. Adare {\it et al.} (PHENIX Collaboration), Phys. Rev. D {\bf 90}, 012006 (2014).

\bibitem{starpi0}
B. I. Abelev {\it et al.} (STAR Collaboration), Phys. Rev. D {\bf 101}, 222001 (2008).

\bibitem{zgs}
R.~D Kelm {\it et al.}, Phys. Rev. Lett. {\bf 36}, 929 (1976).

\bibitem{ags}
C. E. Allgower {\it et al.} (E925 Collaboration), Phys. Rev. D {\bf 65}, 092008 (2002).

\bibitem{romanpot}
S. B$\ddot{\textrm{u}}$ltmann {\it et al.} Nucl. Instrum. Meth. \textbf{535}, 415 (2004).


  




\end{thebibliography}
\end{document}